\documentclass[conference]{IEEEtran}
\IEEEoverridecommandlockouts
\usepackage{url}
\usepackage{cite}
\usepackage{amsmath,amssymb,amsfonts}
\usepackage{algorithmic}
\usepackage{graphicx}
\usepackage{textcomp}
\usepackage{xcolor}
\usepackage{graphicx}
\usepackage[utf8]{inputenc}
\usepackage{enumitem}
\usepackage{listings}
\def\BibTeX{{\rm B\kern-.05em{\sc i\kern-.025em b}\kern-.08em
    T\kern-.1667em\lower.7ex\hbox{E}\kern-.125emX}}
\begin{document}

\title{Extracting Layered Privacy Language \\ Purposes from Web Services}

\author{\IEEEauthorblockN{Kalle Hjerppe}
\IEEEauthorblockA{University of Turku, Finland \\
kphjer@utu.fi}
\and
\IEEEauthorblockN{Jukka Ruohonen}
\IEEEauthorblockA{University of Turku, Finland \\
juanruo@utu.fi}
\and
\IEEEauthorblockN{Ville Leppänen}
\IEEEauthorblockA{University of Turku, Finland \\
ville.leppanen@utu.fi}
}
\maketitle

\begin{abstract}
Web services are important in the processing of personal data in the World Wide Web. In light of recent data protection regulations, this processing raises a question about consent or other basis of legal processing. While a consent must be informed, many web services fail to provide enough information for users to make informed decisions. Privacy policies and privacy languages are one way for addressing this problem; the former document how personal data is processed, while the latter describe this processing formally. In this paper, the so-called Layered Privacy Language (LPL) is coupled with web services in order to express personal data processing with a formal analysis method that seeks to generate the processing purposes for privacy policies. To this end, the paper reviews the background theory as well as proposes a method and a concrete tool. The results are demonstrated with a small case study.
\end{abstract}

\begin{IEEEkeywords}
Data protection, privacy engineering, privacy language, static analysis, semantic web, GDPR, LPL, SOA
\end{IEEEkeywords}

\section{Introduction}

There has been a long-standing gap between research and practice in the domain of privacy and data protection~\cite{martin2018methods}. Privacy engineering has emerged as a field that seeks to narrow this gap. In general, privacy engineering can be defined as a ``\textit{field of research and practice that designs, implements, adapts, and evaluates theories, methods, techniques, and tools to systematically capture and address privacy issues when developing sociotechnical systems}''~\cite{gurses2016privacy}. The results from this kind of research can lower the adoption costs for industry and bridge the gap between practice and research. Within this domain, privacy-friendliness of a system can be understood to range from ``\textit{privacy-by-policy}'' to ``\textit{privacy-by-architecture}''~\cite{spiekermann09}. This paper can be positioned to the middle; the goal is to move toward ``\textit{privacy-by-architecture}'' tenets.

Privacy policies are necessary for many software engineering practitioners. For example, the General Data Protection Regulation (GDPR) requires those processing personal data to keep a record of their data processing activities~\cite{gdpr}. Such bookkeeping necessitates the documentation of the processing activities in privacy policies. Documentation is required also for establishing the lawfulness of a given processing activity. In the Web, the consent of data subjects is presumably the most widely used way to establish the lawful basis of processing personal data under the GDPR. But to obtain an informed and freely given consent, a data subject must know and understand the purpose(s) of processing his or her personal data. Against this backdrop, it becomes understandable why the whole concept of consent has long been recognized as problematic in the digital era \cite{giannopoulou2020algorithmic}. There are many practical reasons behind the problems. For one, there exists an \textit{information asymmetry} between service providers and data subjects~\text{\cite{soh2019privacy,bashir2015online}}. This asymmetry motivates developing new methods for expressing privacy policies in more transparent and understandable ways. At the same time, there is value for service providers in tool-assisted privacy policy management.

As an illustration, consider a web service (WS) that processes personal data in a context of booking a hotel room online. In this context, ``hotel room reservation'' might denote the high-level purpose for the processing. However, this purpose does not provide enough details to obtain an informed consent. Therefore, the purpose could be augmented to contain more fine-grained elements, such as ``authentication'', ``review display'', ``account management', ``reservation confirmation'', and ``online payment''. In a typical WS implementation, these specific elements map to particular software architecture modules in a service-oriented architecture (SOA). By further increasing the granularity and lowering the level of abstraction, it is possible to enumerate these elements as a sequence of WS-specific functional calls. By using a formal privacy policy language, the enumeration can be further used to generate policies that describe the distinct elements in the processing. It should be noted that this example only covers processing done by a primary service provider. In practice, transfers of data to other domains often occur; a single policy does not cover the entire transitive closure in the flow of personal data. A payment transaction might use a third-party provider, for instance. To restrict the scope, this paper only considers first-party processing of personal data, although the theory described does take also third-party transfers into account.

In terms of privacy engineering, the previous illustration underlines the possibility to switch between a \textit{user-focused} perspective of \textit{validating} a privacy policy against a concrete software architecture, and a \textit{developer-oriented} perspective of \textit{generating} the policy automatically~\cite{ifippaper}. A tool that can carry out such switching facilitates the writing of privacy policies---even though human intervention is still required in practice. Given this brief motivation, the present work lays down the groundwork for generative privacy policies. In essence, a privacy policy for a web service is also a documentation of its behaviour. Against this backdrop, the paper's goal is to demonstrate an application programming interface (API) for documentation that fits into a privacy policy. To this end, the following two research questions (RQs) are examined:

\begin{itemize}

\item{RQ1: \textit{How to formalize a model describing personal data processing for web services?}}

\item{RQ2: \textit{How to automatically extract this data from code?}}

\end{itemize}

The answers to these two questions establish both a \textit{method}~(RQ1) and a \textit{tool}~(RQ2). In terms of the former, existing models for privacy policies are extended and augmented in the web application context; in terms of the latter, a concrete implementation is presented. From a practical privacy engineering viewpoint, it is important that the results are possible to integrate into real-world applications with a low amount of costs. Regardless of a particular domain, such low-cost integration is always a challenge for formal languages.

The remainder of the paper is structured as follows. The opening Section~\ref{background} outlines the background and related work. The subsequent Section~\ref{definitions} provides the formal definitions for the method and the tool thereto. These definitions are provided in two parts: the LPL is first extended to accommodate composition, after which the language's $Purposes$ are mapped to web services. Thus, this section provides the answer to RQ1. For answering to RQ2, Section~\ref{implemenetation} presents a concrete implementation based on static analysis and SOAs. The use of the tool implemented is further elaborated in Section~\ref{casestudy} with a small case study. Finally, Section~\ref{discussion} discusses the answers reached and pinpoints a few directions for further work.

\section{Background and related work}\label{background}

Multiple competing formal privacy languages have been presented in recent years. Of these, the LPL is a good and timely example as it has been explicitly designed to address the GDPR's requirements~\cite{lplgerl}. It also adheres to both legal and technical privacy viewpoints. The main design requirements for the language included the differentiation of data subjects and data recipients, purpose-based policies for data, retention and anonymization elements, the ability to layer policies for provenance, and human-readability. While some previous work exists also for programmable privacy languages~\cite{ramezanifarkhani2018secrecy}, the LPL's abstraction level focuses on the modeling of privacy policies, and the level is also suitable for the present work.

After the LPL's initiation in 2018, further effort has been devoted for extending the language and building features around it. The examples include authentication~\cite{wilhelm2019policy}, personalization~\cite{gerl2019let}, and privacy icons~\cite{gerl2018extending}. However, there appears to be no previous work for mapping LPL specifically to web services, nor does there seem to exist previous research for grouping purposes. Although LPL itself features a $PurposeHierarchy$, this feature is an inheritance relationship. In contrast, the present work operates with composition.

One goal of formal languages for expressing privacy policies is to make them machine-readable. This readability enables algorithmic validation of a privacy policy. There exists also some previous work in this regard. For instance, a language has been developed for the GDPR's requirements~\cite{bonatti2020machine}. Compared to the original, unmodified LPL, the language's ontology is more expressive, modelling consent, processing of data, location, and related data protection characteristics. If the goal is compliance checking instead of mere formulation of a policy, such modeling is also necessary. Against this backdrop, the present work builds upon LPL and takes it a step closer to implementation. That said, the concepts presented (excluding Section \ref{lplextension}) are applicable to other policy languages as well.

In essence, the GPDR entails six aspects for privacy policies: the purposes of data processing, the data processed, the potential data recipients, transfers of the data, erasure conditions for the data, and information about the processing itself \cite{bonatti2020machine,lplgerl}. In its current state, LPL covers these aspects, but it is cumbersome to model details about the processing in a transparent way. Therefore, the present work further provides means to document web services in detail and to generate this data automatically. There exists plenty of previous work in the web service context, of course. The so-called semantic web would be a prime example~\cite{tumer, verborgh2014survey}. There are also many industry tools and standards for API documentation, such as the OpenAPI initiative.\footnote{~\url{https://www.openapis.org/}} In general, these API frameworks provide means to describe a service's functionality and to increase its discoverability~\cite{ed2017example}. The OpenAPI specification has also been extended with metadata in order to improve interoperability~\cite{dastgheib2017smartapi,zaveri2017smartapi}. Discoverability and interoperability also align with the GDPR's high-level goals. Modelling privacy requirements by combining the legal view with a technical view using Petri nets also has been presented \cite{diver2017opening}.

Generating API documentation from source code is also an industry standard practice, and there are tools that integrate this task with OpenAPI.\footnote{~\url{https://swagger.io/tools/swagger-codegen/}} In addition to documentation, these generators can create client libraries which use the APIs. Despite these advances, there appears to be no existing tools specifically tailored for generating privacy policies. By building upon an existing solution for annotating personal data processing in source code~\cite{ifippaper}, this paper sets to narrow this apparent gap in both research and practice.

\section{Definitions}\label{definitions}

This section first defines a composed $Purpose$, an extension to the Layered Privacy Language. This definition creates a bridge from the formal language to the technical implementation level. After having demonstrated that a concrete $Purpose$ can be represented as a directed graph, it is subsequently elaborated how these concepts can be mapped to web services.

\subsection{Defining \textit{Purpose} composition}\label{lplextension}

The layered nature of LPL allows for nested privacy $Policies$, and $Purposes$ can have inheritance. While this is useful, our application requires another way to layer granularity and abstraction levels into the language---the component parts that $Purpose$ consists of. Here, \textit{inheritance} is understood as an \textit{is-a} relationship; a child may be substituted for its parent. \textit{Composition}, in turn, refers to a \textit{has-a} relationship of holding a reference to another component element. The LPL forbids multiple inheritance, but composition has no such restriction.

A $LayeredPrivacyPolicy$, or $lpp$, that is, an element representing a privacy policy, is defined \cite[Def.~1]{lplgerl} as a tuple:
\begin{equation}\label{eq: lpp}
lpp = (version, name, lang, ppURI, upp, \widehat{P}),
\end{equation}
where $version$ describes the LPL version, $name$ labels the given policy, $lang$ specifies the language with which the policy is written, $ppURI$ is a link to a textual version of the policy, $upp$ is a so-called $UnderlyingPrivacyPolicy$ element, and $\widehat{P}$ is the set of $Purposes$ the policy consists of. Of these, $upp$ allows composing policies from different entities, while $\widehat{P}$ defines the actual policy content. An LPL $Purpose$ element, $p$, is defined \cite[Def.~6]{lplgerl} as the following tuple of values:
\begin{equation}\label{eq: purpose}
p = (name, optOut, required, descr, \widehat{DR}, r, pm, \widehat{D}),
\end{equation}
where $\widehat{D}$ is a set of $Data$ elements accessed in the given $Purpose$, $\widehat{DR}$ a set of $DataRecipients$, $r$ a $Retention$ element, and $pm$ a $PrivacyModel$ element. The boolean values $optOut$ and $required$ define whether the purpose has to be ``actively denied'' and whether the purpose has to be accepted to be able to consent to the policy. Finally, $name$ and $descr$ are human-readable textual descriptions.

Given these preliminaries, the goal is to extend the language by defining the relation among composite $Purposes$ in a $Policy$. Since a root-level $Purpose$ in a privacy policy is often not specific or concrete enough to map to a technical implementation unit, the hierarchy between $Purposes$ can be defined as a directed graph. However, abstract concept-level purposes (e.g., ``reserve a room'') contain many concrete parts (e.g., API calls). Therefore, the aim is to also define constraints under which data subjects accepting a given $p$ may accept all its composite purposes, denoted by $p'$, without weakening their data protection. The relation is specific to a $Policy$, which defines the hierarchy---policies may have different hierarchies. 

Let $up$ denote a relation for a $UnderlyingPurpose$, represented as a tuple of values in a policy, and $\widehat{up}$ a set of $up$. Further denote the relation of all $ComposedPurposes$ in an $lpp$ by $cp$. Then,
\begin{equation}\label{eq: up cp}
up = (p, p')
\quad\textmd{and}\quad
cp = (lpp, \widehat{up}) ,
\end{equation}
where $p$ is any $Purpose$ in the policy, and $p'$ is a $Purpose$ element with additional constraints to form a valid privacy policy, as follows. From the relationship that a composite $p'$ is a component of a $Purpose$ $p$, the constraints for a $Policy$ can be reasoned. In any given $lpp$, where $p$ is a $Purpose$ belonging to it and $p'$ its underlying $Purpose$, i.e.~the tuples in \eqref{eq: up cp} exist and $up \in \widehat{up}$, in order for the $lpp$ to be valid, it must hold that
\begin{equation}\label{eq: D}
\widehat{D}_{p'}  \subseteq  \widehat{D}_{p} ,
\end{equation}
\begin{equation}\label{eq: DR}
\widehat{DR}_{p'}  \subseteq  \widehat{DR}_{p} ,
\end{equation}
\begin{equation}\label{eq: R}
r_{p'}  \leq  r_{p} ,
\end{equation}
\begin{equation}\label{eq: Pm}
pm_{p'}  \geq  pm_{p} ,
\end{equation}
\begin{equation}\label{eq: required}
required_{p'}  =  required_{p} ,
\end{equation}
\begin{equation}\label{eq: Opt}
optOut_{p'}  =  optOut_{p} .
\end{equation}

In \eqref{eq: R} the inequality is defined as a strictness of a \textit{Retention} element depending on its particular $type$. Let $Indefinite \geq AfterPurpose \geq Fixed\-Date$ and assume that comparisons with the same $type$ are decided by $pointInTime$. With these assumptions, $AfterPurpose \geq FixedDate$ follows from not considering a $Purpose$ completed until all $FixedDate$ retentions are resolved. It is also worth remarking that the comparison of $PrivacyModel$ elements in \eqref{eq: Pm} is not defined strictly. The reason originates from the LPL, which does not provide a rigid definition for a set of potential (pseudo)anonymization methods. A comparison of two $pm$'s with the same $name$ (e.g., ``\textit{$k$-anonymity}'') is decided by the values of their $PrivacyModelAttributes$. Without a definition of an exact comparison across different $pm$ types, it is possible to avoid the issue by augmenting the criteria with a requirement that the $type$s of $pm_{p'}$ and $pm_{p}$ must be~equal.

The definition for $ComposedPurposes$, $cp$, accompanies another structure originally defined in the LPL that is named $PurposeHierarchy$, $ph$. $PurposeHierarchy$ defines a parent-child relationship as an inheritance hierarchy. While this hierarchical \textit{is-a} relationship is similar to the \textit{has-many} relationship for $ComposedPurposes$, there exists a meaningful difference in capability and intent. In other words, the relationships are complementary. Whereas inheritance enables reusing rights and making a $Purpose$ concrete, $ComposedPurposes$ expose the actual contents of the $Purpose$. Given that an inherited $Purpose$ is substitutable to its parent, it cannot elaborate its intent to a data subject. A brief example can be used to elaborate this point further. 

\begin{figure}
\centering
\includegraphics[width=\linewidth]{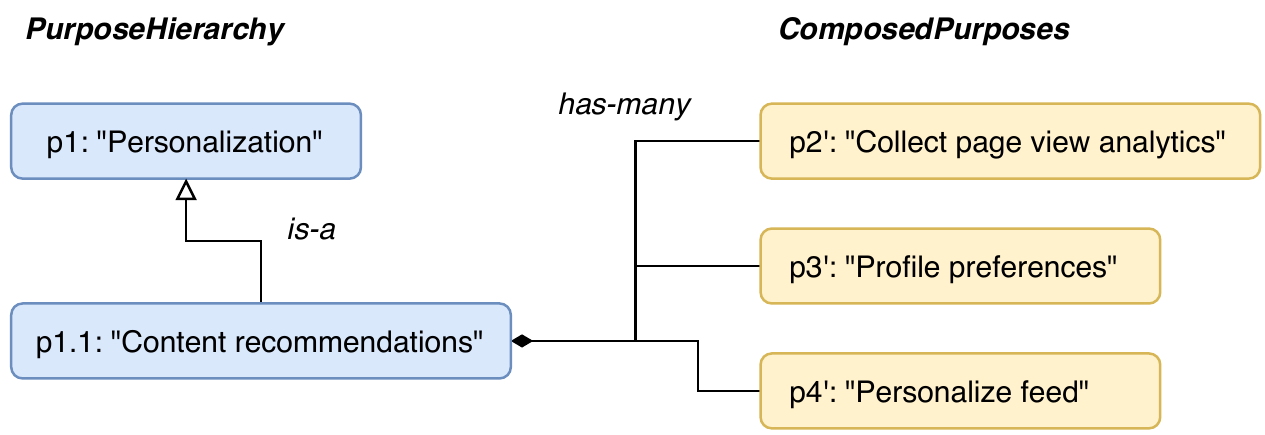}
\caption{An example of a possible \textit{PurposeHierarchy} using \textit{ComposedPurposes}, highlighting both composition and inheritance.}
\label{personalization}
\end{figure}

Thus, consider ``personalization'' as a parent $Purpose$. This purpose, $p1$, might be inherited by ``content recommendations'', $p1.1$. Both would inform the data subject about the categories of personal data used, but these would not expose any further insights. But by adding underlying purposes to $p1.1$, it is possible to elaborate the nature of ``content recommendations'' further. For instance, in Fig.~\ref{personalization} the purpose $p1.1$ composes ``collect page view analytics'' ($p2'$), ``profile preferences'' ($p3'$), and ``personalize feeds'' ($p4'$). Clearly, when compared to displaying only $p1.1$, these compositions make the policy more transparent to the data~subject.

It is possible to construct a policy that does not meet the properties required, but it would not be \textit{valid}. Using the definition of $up$ in \eqref{eq: up cp}, it is possible to display a privacy policy organized as a directed acyclic graph. A cyclic graph could be theoretically viable in a policy (as long as the constraints for valid policies are met), but we have not found meaningful use-cases. The root elements are those which are not underlying any other purpose---and following the constraints, the data subject needs only to study and accept these. An interested party may read deeper. Furthermore, the formal definitions enable the subsequent functionality: coupling the purposes with technical~functions.

\subsection{Web services as privacy policy purposes}\label{wsappp}

The composed purpose specifications elaborated allow to create more informative privacy policies. These empower to specify down from an abstract level of a personal data processing \textit{Purpose} to what the processing concretely consists of. These specifications enable a deeper validation of a policy, more transparency, and confidence that the policy matches a corresponding technical implementation. Although the composed $Purpose$ is entirely domain-agnostic, it is particularly useful in the web application domain.

Therefore, consider a web service as a function of requests to responses (see the definition of a server in RFC 2616; \cite{fielding1999rfc}). To increase the abstraction depth, these basic characteristics can be further extended with the concepts of web services and service nets (SNs) \cite{hamadi2003petri}. Roughly, a web service is a tuple of values describing the service, a set of component web services, and a service net. An SN, in turn, is a place-transition Petri net modeling the dynamic behavior of a given web service. To proceed more formally, let
\begin{equation}\label{eq: ws}
WS = (NameS, Desc, Loc, URL, CS, SN),
\end{equation}
where $NameS$ is the service's unique identification code, $Desc$ a textual description for it, $Loc$ the server in which the service is located, $URL$ its endpoint, $CS$ a set of component web services, and $SN$ the service net describing its behaviour~\cite{hamadi2003petri}. A service net, in turn, can be defined as a~tuple
\begin{equation}\label{eq: sn}
SN = (P, T, W, i, o, \ell),
\end{equation}
where $P$ is a set of \textit{places}, $T$ a set of \textit{transitions} representing the operations of the service, $W$ a set of directed \textit{arcs}, $i$ the input place, $o$ the output place, and $\ell$ a labeling function for transitions~\cite{hamadi2003petri}. A couple of assumptions allow to connect these definitions to personal data processing. First, in the present context, any web service is assumed to be governed by a privacy policy, and that policy can be modelled with LPL. Second, the dynamic behavior exposed by the Petri net model can be exploited by assuming that personal data processing occurs always within $T$. The latter assumption necessitates a more thorough inspection of the behavior of a service, and assumes it is possible to do so.

For a simple instance, say that an $SN$ of a web service has an input place $i$, an output $o$, and three transitions with labels  ``\textit{register customer}'' ($t1$),  ``\textit{create subscription}'' ($t2$), and  ``\textit{send confirmation email}'' ($t3$). The set of arcs $W$ connect $i \to t1 \to t2  \to t3 \to o$. Each transition processes its own set of personal data. If this processing can be inspected, a union can be used to define the total processing for the given $SN$. The principle remains the same in a more complex instance with branches: a union is the sum of all potential transitions in the service~net.

To this end, a function that inspects personal data processing in any given web service can be defined as:
\begin{equation}
pd: WS \to (\widehat{D}_{ws}, \widehat{DR}_{ws}, \widehat{UPP}_{ws}),
\end{equation}
where $WS$ is a web service and the return value is a tuple describing data processing in its behaviour. In this tuple, $\widehat{D}_{ws}$ is a set of personal data (i.e., LPL's $Data$ elements) processed in the transitions of a service net of the $WS$, $\widehat{DR}_{ws}$ is a set of authorized parties allowed to access the personal data (i.e., LPL's $DataRecipient$ elements), and, finally, $\widehat{UPP}_{ws}$ is a set of privacy policies (i.e., $UnderlyingPrivacyPolicy$ elements) of third-parties the WS transfers personal data to.

The function $pd$ allows to model personal data processing at a sufficiently abstract level. Also the mapping of the definition \eqref{eq: ws} to LPL $Purpose$ is straightforward by using the composition rules described earlier. In essence: (i)~any $WS$ processing personal data is governed by a privacy policy; (ii)~the act of \textit{processing} in LPL is encoded in $Purposes$; and, therefore, (iii)~at least one $Purpose$ governs any web service processing personal data. In other words, the definitions \eqref{eq: purpose} and \eqref{eq: ws} are composable and flow from a high abstraction level to concrete operations. These can be further mapped to the definition \eqref{eq: lpp} by noting that (iv)~there exists a $Purpose$ for any web service $WS$ in a set of web services $\widehat{WS}$ governed by a \textit{LayeredPrivacyPolicy}. In other words:
\begin{equation}
    \textmd{for all}~WS \in \widehat{WS}~ \textmd{exists}~p \in lpp
\end{equation}
for which
\begin{align}
pd(WS) := (
&\widehat{D}_{ws} \subseteq \widehat{D}, \notag \\
&\widehat{DR}_{ws} \subseteq \widehat{DR}, \\ \notag
&\widehat{UPP}_{ws} \in lpp) .
\end{align}

Finally, in order to simplify the notation, let the following tuple mark the relation between a $Purpose$ and a web service:
\begin{equation}
    gov = (WS, p) .
\end{equation}
Although the relation $gov$ is not specified as a part of a privacy policy, this mapping between the policy and a description of a service provides a useful method for data controllers. Although any web service processing personal data does so under a privacy policy, this condition does not dictate the practice. For instance, a single $Purpose$ could govern all web services of a policy, or a single $WS$ might be governed by multiple $Purposes$. In practice, it is also possible that a policy is \textit{invalid}: a $pd(WS)$ does not match any $Purpose$ in a policy. Such cases should be avoided, obviously.

\begin{figure}
\centering
\includegraphics[width=\linewidth]{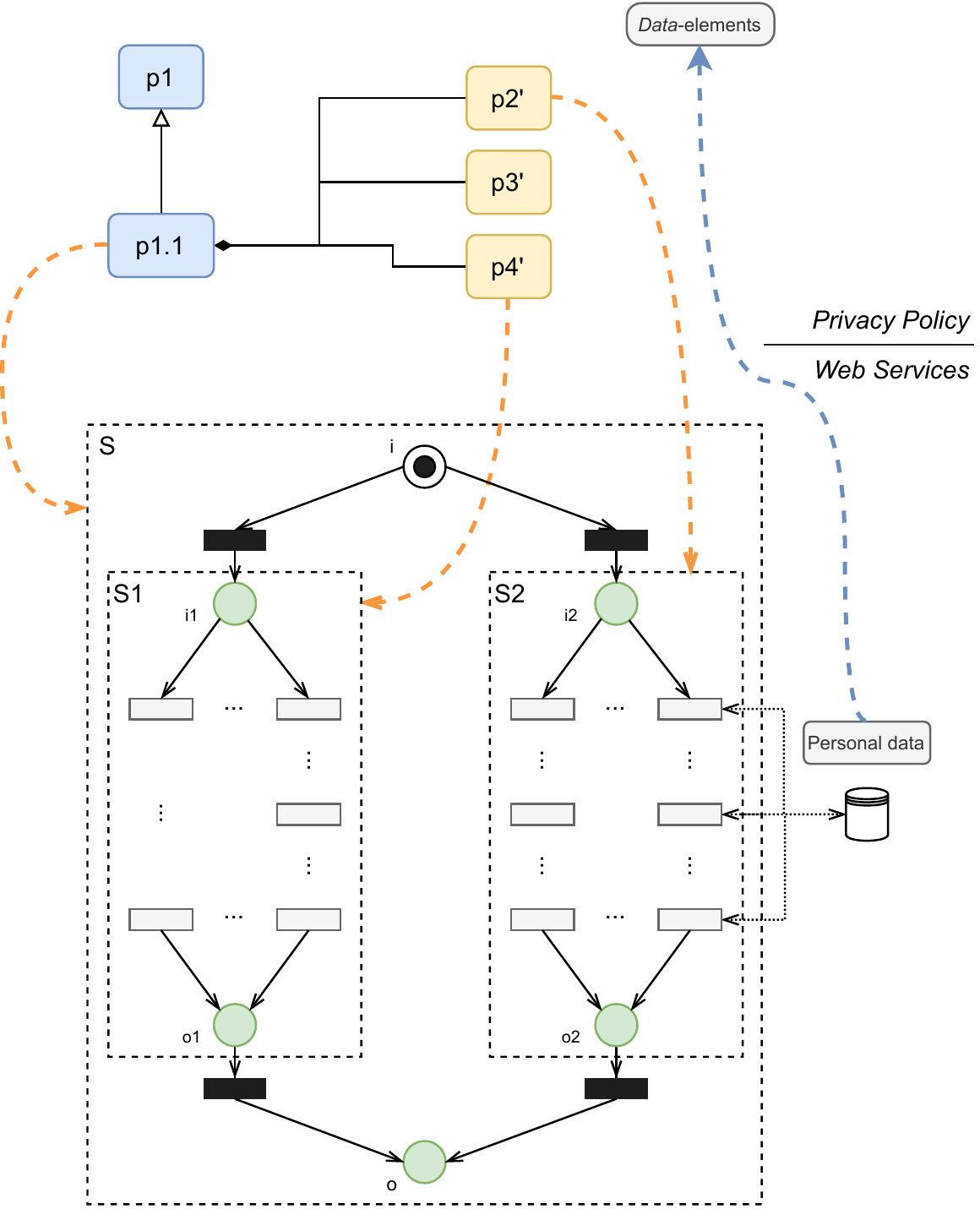}
\caption{Illustration of LPL $Purposes$ mapping to composed web services; dotted arrows represent the relation $gov$ and the function $pd$ (extended from \cite{hamadi2003petri})}
\label{lplws}
\end{figure}

A further benefit originates from the fact that modeling can be done at either side of the abstraction; at the level of web services or at the level of privacy policies. It is possible to ``attach'' a governed web service at any level of the LPL \textit{Purpose} graph. Also composed web services can be modeled. To illustrate these points, Fig.~\ref{lplws} shows a web service $S$ composed of two services $S1$ and $S2$, both of which have a governing $Purpose$. The $Data$ elements are derived from the $SN$ of the web service in question through the function $pd$. The depth and granularity of a privacy policy can be chosen as it is written. Different configurations and their privacy implications are discussed later on in Section~\ref{discussion}.

\section{Implementation}\label{implemenetation}

\subsection{Static analysis}\label{staticanalysis}

Since the personal data processing purposes coupled to web services are derived from the behaviour of the services, the next part is to define a way to generate the data via static analysis. This construction requires limiting the scope to the specific implementation level. To still retain a high degree of generality, the practical solution discussed focuses on a particular but common design pattern for SOA web services. 

In essence, many web frameworks use the same strategy of defining application entry-points in the source code by using annotations. Good examples include the Spring framework\footnote{~\url{https://spring.io/}} for Java and Flask\footnote{~\url{https://pypi.org/project/Flask/}} for Python. In this pattern, the server functionality is run by the given library, which maps incoming requests to (application code) function calls based on annotated end-point definitions. Strictly speaking, the pattern is a composite web service, which conditionally calls a different component web service based on the given request. Considering it as a single service with multiple entry-points is a reasonable shorthand; the source code for many entry-points may belong to a single file or class, for instance.

In the present work, each entry-point maps to a privacy policy \textit{Purpose}, as defined in the previous Section~\ref{wsappp}. By continuing the annotation-based approach, the $Data$ elements can be derived from \textit{object relational mapping} (ORM) classes. Many web service frameworks offer annotations to mark these declaratively and to define the relational mapping statically (see Hibernate\footnote{~\url{https://hibernate.org/orm/}}, for instance). Personal data stored in these database entity objects can be reasoned about if the semantic information is provided in a similar way \cite{ifippaper}.

Using both the application entry-point and personal data \textit{ORM} annotations, a code-base declares how personal data is processed statically. This declaration maps to the theoretical service net definition in a concrete manner. By using static analysis processing, it is possible to create (at the least templates for) \textit{Purpose} definitions. Since each entry-point is a ``main'' function of a subprogram of a web service, they form a distinct directed graph of function dependencies. Analyzing each function of the code-base during compilation allows marking personal data the function depends on. To demonstrate this processing, a concrete implementation was developed for the Java programming language.

The method and the implementation address generating $Data$ elements from source code in a specific architecture. Essentially, the static analysis process represents an implementation of a part the $pd$ function. Full capabilities of the inspection function $pd$ would require providing the $DataRecipients$ the web service has as well as the set of $UnderlyingPrivacyPolicies$ of third-parties the $WS$ transfers data to. Although both requirements are possible to fulfill, these require heavier contextual data to be present in the code. Storing this data in annotations may not be the optimal way. Therefore, these are excluded from the scope of the present work. Another point is that extracting $Data$ elements from source code is not enough generate complete $Purposes$. While that might be the end goal, even incomplete purposes are useful on their own right. For instance, the values from the tuple $p$ in \eqref{eq: purpose}, including $name$ and $descr$, can also be generated with static analysis and matching annotations. That said, the values for $optOut$, $required$, retention $r$, and privacy model $pm$ require human judgement in the present work.

\subsection{Data extraction tool}

A concrete implementation was developed for the method described. The data extraction part of the program was developed as a compiler plugin for the Java language. The tool runs as an annotation processing tool (APT) and generates a fresh dataset automatically upon building a project it is enabled on. The source code of the tool is published under an open source license.\footnote{~\url{https://github.com/devgeniem/personaldataflow}} Using the APT interface, the package can be integrated to both command line tools and integrated development environments. 

The prototype implementation is limited to a particular set of supported annotations using the Spring framework: \texttt{@RequestMapping} and \texttt{@Document}. In addition, the implementation requires the use of the personal data annotation \texttt{@PersonalData} described in previous work~\cite{ifippaper}. It should be relatively trivial to extend the support for more framework annotations as long as the target architecture is similar. Extending the method to further programming languages is more challenging, but not impossible. As of yet, only the data extraction part is fully automatic: visualizing the results in an IDE plugin is a possible opportunity for future work.

Developing the inspection process for the specific environment required some further design choices. Given that there are no right or wrong answers in this context, other implementations might differ slightly. In particular, Java's Interfaces are handled with a pessimistic approach. For instance, consider a $Purpose$ entry-point using an interface with multiple implementations. The tool implemented handles this by summing the personal data found in all available implementations. Another design choice made pertains to other modules (libraries) imported in a web service. As the tool is tied to compile-time static analysis, other (pre-compiled) modules are out of the scope of the tool. In practice, this choice was not seen as a hindrance because the given business logic generally defines the type of personal data processed. Finally, access to client-side libraries of third-party services would be important when analysing \textit{transfers} of personal data.

In summary, the tool presented satisfies the data extraction for a specific case of the more general solution outlined. The concrete implementation is limited to a single architecture, but extending to different design patterns is possible in the future. The subsequent section briefly describes how the static analysis tool was used in practice.

\section{A case study}\label{casestudy}

To demonstrate the practical use of the tool implemented, the tool was applied to an industry code-base of a web service. The goal of the case study was to validate the tool with a simple hypothesis in mind: the generated documentation should match the theoretical expectations.

The case WS has been in production for processing transactions in a Finnish company since 2014, under different maintainers and product owners. Despite ongoing maintenance and development, there was no automated documentation until the deployment of the tool presented. Deploying the tool required annotating the personal data database entity classes and installing the annotation processor to the build chain. This integration amounted to a moderate amount of work. The moderate work amount indicates applicability of the tool to new targets with roughly similar software architectures. 

The case $WS$ is a single monolithic code-base, which is split into modules. The total of 323 classes (amounting to about 28 thousand lines of code) can be understood as a \textit{Model/View/Controller} architecture with specific service and database layers. Considering the definition for composable web services, the entire system would be the root $WS$. It would compose $WS$ modules (``Controllers''), which, in turn, compose $WS$ endpoint functions. These functions are governed by composed $Purposes$, which the data extraction tool maps. Although the case $WS$ uses several libraries, these are not relevant for the case study. It should also be noted that authorization for the end-points is out of scope; the sessions are checked in the framework rather than in the code inspected.

The results of the data extraction tool can be summarized as follows. The case $WS$ was found to compose of 30 Controller modules, each of which compose of multiple end-points (a total of $245$ with a range of $[1, 51]$, and an average of $8.2$). Out of these, not all processed personal data. After filtering out those, a total of $22$ Controller modules remained (with a total of $224$, a range of $[1, 51]$, and an average $9.3$). In total, $19$ different personal data entity types were identified.

\begin{figure}
\centering
\includegraphics[width=\linewidth]{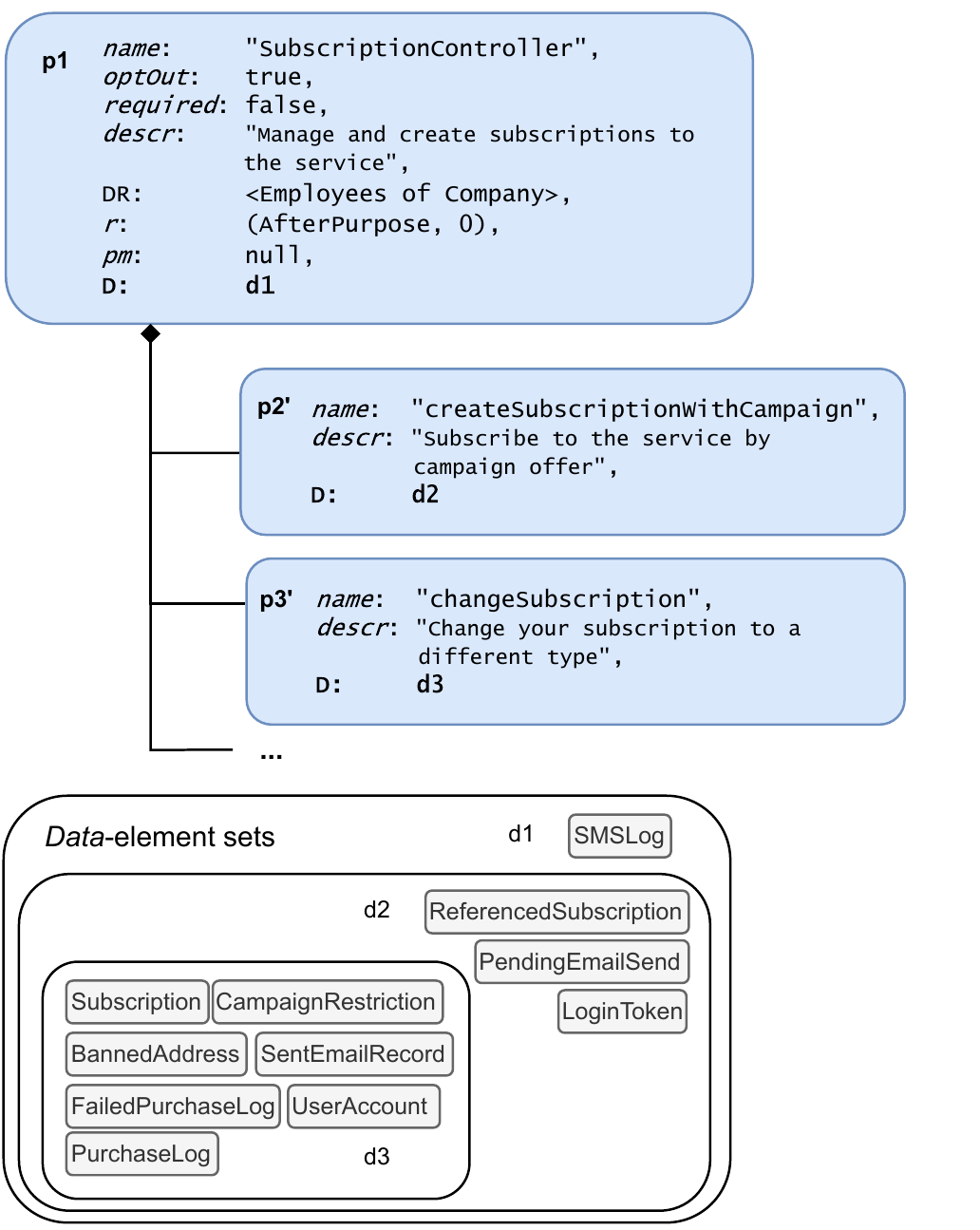}
\caption{An example of composed $Purposes$ with their corresponding $Data$-elements from the case study.}
\label{case}
\end{figure}

Following our formulation of the problem, the case $WS$ was viewed as set of component web services (i.e., Controller modules), which in turn are composed of entry-point web services. An example of the data extracted via static analysis is visualized in Fig.~\ref{case}, where the resulting component of the case $WS$ are accompanied by the sets of personal data processed within the $Purpose$. The analysis tool also calculates the same data for the entry-point $WS$, which is omitted in this figure. As defined before, the sets of $Data$ elements belonging to any component $Purposes$ $p'$ are subsets of their higher-level $Purpose$ $p$. The data also visualizes the application architecture, and highlights important areas; those where many entities concentrate in a single $Purpose$ as well those where entities are used across multiple purposes.

\begin{figure}
\centering
\includegraphics{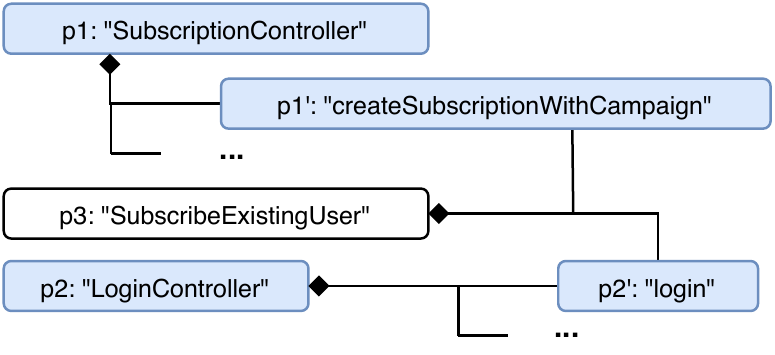}
\caption{Another example of composed $Purposes$ generated in the case study, with an added missing $Purpose$ (white) inferred from client-side.}
\label{case_missing}
\end{figure}

There is no existing formal privacy policy for the case $WS$, so an exact number for coverage cannot be given. Regardless, it is possible to gain insights by comparing the implicit $Purposes$ in the system to those extracted with the tool implemented. With manual verification, it is confirmed that the results of the analysis cover all of the entry-points. Thus, in this sense, the tool is sound. However, the complete logical hierarchy of the composed $Purposes$ cannot be constructed by inspecting the $WS$ source code alone. As a counterexample, the process of purchasing (a purpose) composes of entry-point $Purposes$ from $WS$'s $SubscriptionController$ and $LoginController$. This can only be inferred by viewing the clients in addition to the web Services. An example of this case is illustrated in Fig.~\ref{case_missing}, where composed $Purposes$ $(p1, p1')$ and $(p2, p2')$ also have a third $Purpose$ $p3$ that could not be found in the web service analysis.

The case study demonstrates the possibility to construct any $Purpose$ concerning the case system, by composing the extracted $Purposes$ that act as building blocks. The presented method thus answers to RQ2. To create the whole $Policy$, it is not sufficient to analyse server-side code alone.

\section{Discussion}\label{discussion}

It is trivial to say that any web service processing personal data works under a privacy policy, whether implicit or explicit. The definitions of this paper only provide the means to express and reason about the processing formally. However, whether the relation $rl$ and the $PurposeHierarchy$ defined are used to actually increase transparency to the data subject is not \textit{enforced} with the results of this paper. The method of automatically generating $Purposes$ from a web service source code might lower the effort required to create more expressive privacy policies. This assumes a good faith effort on part of the service provider: a more detailed privacy policy might be also a liability for maintenance. The first principle of the GPDR requires personal data processing to be lawful, fair, and transparent \cite{gdpr}. In a practical environment, there are multiple other constraints and requirements competing with these abstract principles \cite{hjerppe2019general}. In the end, it is on the shoulders of academic and industry organizations to proliferate these privacy engineering methods to wider adoption via examples and standards. Needless to say, the task is not easy. But by using automated documentation methods, such as the present work, may provide means to move forward by focusing on data controllers instead of data subjects. For a plausible route for adoption with less friction, the documentation generated could be integrated first to internal documentation repositories. From there, it could be further integrated to public privacy policies.

As was remarked in Section \ref{casestudy}, the automatically generated $Purposes$ cannot describe the full privacy policy in most cases. Although it depends on a given system, rarely will the logical processes (purposes) exhaustively map to a hierarchy derived from a service structure. However, it can be claimed that any policy which uses a web service ought to include the $Purpose$ governing that $WS$ in its composition. There should be also one $Purpose$ that governs all web services of a privacy policy. Given these assumptions, a preliminary metric can be proposed for the \textit{transparency} of a privacy policy. Namely: the ratio of $Purposes$ to web services should be more than one as long as the breadth of the $ComposedPurposes$ graph is minimized (i.e., consent is not asked for unnecessary purposes). Although transparency of a privacy policy is a part of its quality, it is difficult to say anything definite about potential quality improvements in general.

For instance, it cannot be conjectured that more information would always imply ``better'' privacy policies. A privacy policy that overwhelms a data subject with information might not be presented in ``\textit{intelligible and easily accessible form, using clear and plain language}'' \cite{gdpr}. In other words, the legal requirements for a consent might not be satisfied. The composed purpose definition of this paper somewhat alleviates this issue: it is possible to specify purposes in detail, and only show deeper levels of a given tree to those data subjects who are interested in the details. To overcome potentially confusing and ambiguous linguistic expressions, more standardized terms and expressions could be used for $Purpose$ descriptions. Also graphical user interface elements and other visual cues could be used to improve the presentation. 

\subsection{Limitations and further work}

The presented work is by no means exhaustive. Leveraging static analysis for privacy policies has further potential. Some limitations of this paper are a deliberate scoping choice, which can be briefly discussed further. The implementation of $pd$ within the constraints defined in Section~\ref{implemenetation} is just one instance of the general problem. However, similar definitions should be possible for other architectures and programming languages.

The inspection function $pd$, as defined in this paper, only handles the first aspect of it: finding personal data processed. Finding \textit{transfers}of data to realms of other privacy policies would be an important step towards computer-assisted privacy policies. A simple approach would be to encode also these with annotations (i.e., a web service code executes a transfer). Another possibility would be to leverage further semantic web capabilities by finding the underlying policies from the source. 

As discussed previously, the framework of generating $Purposes$ from web services does not have the full information for a complete $Policy$ on its own. In addition to the $Purposes$ matching web services, a complete $ComposedPurposes$ graph requires client side information to know the combinations on how the building blocks are arranged. Further research might analyse client-side source code in order to combine data with the web service $Purposes$. Although a similar approach would suffice, the method would require novelty; if web services are understood as a tree, client-side code would be a pyramid, in a manner of speaking.

The case study presented in this work should generally be considered as a validation project. While an example was presented on how the method works, a more comprehensive study might be useful. For instance, a better theory around what constitutes a ``good'' arrangement of $Purposes$ to web services could be studied. Another path forward would be to integrate LPL (or another related language) to the OpenAPI standard (or a similar specification), and generate the corresponding documents automatically. This path would fully start to leverage formal privacy languages in the semantic web.

\subsection{Conclusion}

This paper sought answers to two research questions. The first asked about a way to formalize a model describing personal data processing in web services. To this end, the Layered Privacy Language was extended to include composition for $Purposes$. This extension was formally combined with the definition of web services to form a final model that couples privacy policies web services together.

The second question solicited a method for extracting the corresponding data automatically from a web service code base. This question was approached both theoretically and practically. In theory: by defining an inspection function that would satisfy $LPL$ requirements of a $Purpose$. In practice: by limiting the scope to a certain web service design pattern and by presenting an analysis tool.

\section*{Acknowledgements}

This research was partially supported by both the Academy of Finland (grant number 327391) and Geniem Oy.

\bibliographystyle{IEEEtran}

\end{document}